\documentclass[doubleblind, conference]{IEEEtran}

\usepackage{latexsym}
\usepackage{amssymb}
\usepackage{wrapfig}
\usepackage{cuted}
\usepackage{caption}
\usepackage{amsmath}
\usepackage{amsthm}
\usepackage{booktabs}
\usepackage{graphicx}
\usepackage{color}
\usepackage{color}
\usepackage{hyperref}
\usepackage{tikz}
\usepackage{caption}
\usepackage{todonotes}
\usepackage{paralist}

\usepackage{multirow} 
\usepackage{subcaption}
\usepackage{algorithm}
\usepackage{algpseudocode}

\newcommand{\BibTeX}{B\kern-.05em{\sc i\kern-.025em b}\kern-.08em\TeX}

\begin{document}
\bstctlcite{IEEEexample:BSTcontrol}



\title{In Trust We Survive: Emergent Trust Learning}


\author{
\IEEEauthorblockN{
Qianpu Chen\IEEEauthorrefmark{1},
Giulio Barbero\IEEEauthorrefmark{1},
Mike Preuss\IEEEauthorrefmark{1},
Derya Soydaner\IEEEauthorrefmark{1}}
\IEEEauthorblockA{\IEEEauthorrefmark{1}
LIACS, Leiden Institute of Advanced Computer Science, Universiteit Leiden, The Netherlands\\
Emails: [q.chen.15, g.barbero, m.preuss,  d.soydaner]@liacs.leidenuniv.nl}
}

\maketitle

\begin{strip}
    \centering
    \includegraphics[width=0.9\textwidth]{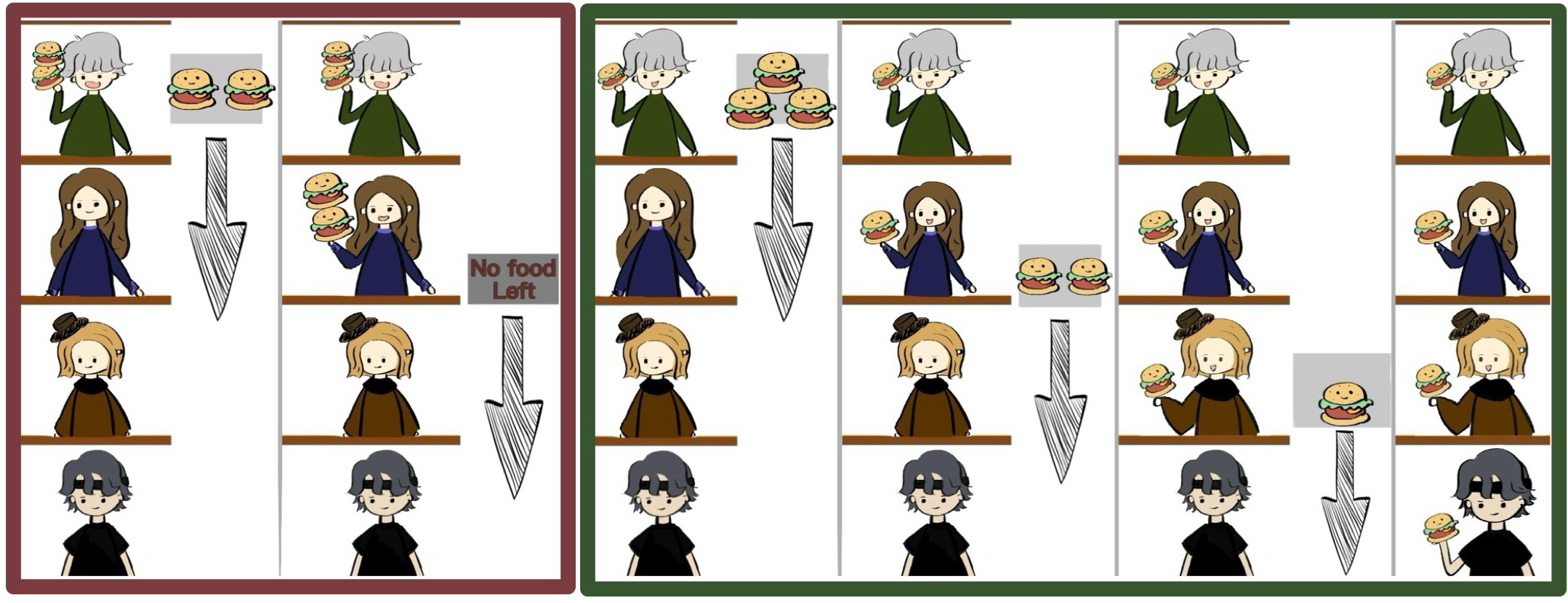}
    \captionof{figure}{Overview of the Tower environment and Emergent Trust Learning (ETL). A food platform starts at the top of a vertical tower and moves down through the floors, each occupied by one learning agent (randomly placed for every run). On each floor, the current agent may eat part or all of the remaining food, which directly improves its own short term reward but leaves less for agents on lower floors. If agents act too greedy, those on upper floors consume too much and agents below eventually starve. As there are no explicit rewards fostering agent cooperation, it has to emerge from behaviour. ETL equips each agent with an internal trust state, memory of recent outcomes, and an adaptive exploration rule based only on local information, thus agents gradually learn to moderate their consumption and maintain a more cooperative and survivable system.}
    \label{fig:tower-overview}
\end{strip}

\begin{abstract}

We introduce Emergent Trust Learning (ETL), a lightweight, trust-based control algorithm that can be plugged into existing AI agents. It enables these to reach cooperation in competitive game environments under shared resources. Each agent maintains a compact internal trust state, which modulates memory, exploration, and action selection. ETL requires only individual rewards and local observations and incurs negligible computational and communication overhead.

\indent We evaluate ETL in three environments: In a grid-based resource world, trust-based agents reduce conflicts and prevent long-term resource depletion while achieving competitive individual returns. In a hierarchical Tower environment with strong social dilemmas and randomised floor assignments, ETL sustains high survival rates and recovers cooperation even after extended phases of enforced greed. In the Iterated Prisoner’s Dilemma, the algorithm generalises to a strategic meta-game, maintaining cooperation with reciprocal opponents while avoiding long-term exploitation by defectors. 
Code will be released upon publication.
\end{abstract}

\begin{IEEEkeywords}
multi-agent reinforcement learning, emergent cooperation, game AI, social dilemmas.
\end{IEEEkeywords}



\section{Introduction}


In the time of more and more sophisticated AI agents (such as LLMs), collaboration in competitive environments is going to be a major research area, especially in the field of games for the foreseeable future. Collaboration needs to happen between AI agents of different origins as well as between AI agents and humans. This comes with previously unseen problems as shown, for example, in the experiences with Open AI Five \cite{OpenAI5} (team of bots playing Dota~2)
which demonstrated that the collaboration of the trained bots with humans is much harder than expected. The problem of Human-AI interaction is investigated more in recent research on computer games \cite{Zhang2025}. We seemingly need tools that enable agents to learn how cooperative or competitive they shall behave.
In competitive multiplayer games, selfish behaviour is often locally optimal but harmful to long-term system health, for example, repeatedly farming the same resource spawn or exploiting map hotspots makes matches increasingly unfair and predictable, eventually depleting resources and collapsing competition into frustration.
Game designers rarely reward altruism directly; instead, they introduce governance mechanics (temporary lockouts, diminishing returns, region-based restrictions) that shape behaviour without changing win conditions.
From the perspective of learning agents, success requires adapting to these implicit rules to keep the game stable and playable.
It may be desirable to have a plugin component that can help agents change their behaviour towards more collaboration when advisable for their own success, without changing their inner logic. With this work, we provide such a component.

This work asks whether similar stabilising effects can be obtained in Multi-agent Reinforcement Learning (MARL)  without explicit cooperative rewards and without handcrafted shaping terms. 
We approach this question by treating cooperation as an emergent response to game rules rather than an explicit objective.
Instead of modifying the reward function, we equip agents with an internal notion of trust that alters how they interpret and react to environmental outcomes. 

We propose Emergent Trust Learning (ETL), a trust-based learning algorithm for competitive settings with purely individual rewards.
Each agent maintains a compact trust state that summarises how supportive or hostile its recent interactions are.
This state is updated from observable signals such as received resources, survival outcomes, or realised payoffs.
High trust encourages more moderate and coordinated behaviour, while low trust shifts agents toward defensive or self-protective actions.
We do not introduce any explicit cooperative bonuses.
Cooperation must emerge because it helps agents avoid unstable or punishing regimes imposed by game dynamics, not because the reward function encodes altruism.

We study ETL across three environments that form a progression from an abstract matrix game to more concrete game-like scenarios. First, we introduce a grid-based resource game where multiple agents compete over shared renewable resources; local over-exploitation reduces future gains, creating a soft balancing mechanic without explicit cooperation rewards.
Second, the Tower Environment, inspired by the film \emph{The Platform}~\cite{theplatform2019}, an extreme social dilemma in which agents on upper floors can easily starve those below by acting greedily. Third, the Iterated Prisoner’s Dilemma (IPD), a classical abstraction for studying cooperation and defection beyond any specific game world.
Across all environments, we compare ETL with standard tabular baselines.
Trust-based agents consistently keep the system in a healthier regime, trigger fewer conflicts over shared resources, 
prevent permanent depletion, and distribute returns more evenly while achieving comparable or higher individual scores.
In the 
Tower and IPD environments, the same trust algorithm remains effective against both cooperative and exploitative opponents, demonstrating robustness beyond a single game design.

\textbf{Our contributions are as follows:}
\begin{compactitem}
    \item We introduce Emergent Trust Learning (ETL), a lightweight trust-based control algorithm that augments standard value- and policy-based learners with a small number of scalar trust and memory variables while relying only on local observations and individual rewards.   
    \item We present the Tower environment, a hierarchical social dilemma testbed where randomly assigned upper-floor agents can starve those below, providing a reusable benchmark for asymmetric roles and cooperation without explicit cooperative rewards.   
    \item We demonstrate across three 
    environments that ETL steers purely self-interested agents toward 
    stable and cooperative regimes by reducing resource conflicts, maintaining high survival in social dilemmas, and supporting mutually beneficial outcomes in strategic interactions.
\end{compactitem}

The paper is structured as follows: in Section~\ref{sec:use-cases}, we enlist some practical, though hypothetical, use cases for applying the ETL. In Section~\ref{sec:related-work}, we review the state of literature on trust-based systems in game AI and multi-agent learning. Section~\ref{sec:etl} introduces the ETL algorithm in detail. We present the experimental environments and results in Section~\ref{sec:experiments}, followed by a discussion and conclusion in Sections~\ref{discussion} and~\ref{sec:conclusion}.

\section{Envisioned Game AI Use-Cases}
\label{sec:use-cases}

We provide a number of example use cases on how the ETL algorithm can be directly applied in computer game contexts:
\vspace*{1ex}

\begin{compactitem}

\item \textbf{Resource governance/directors}: Game designers attach ETL-based trust control 
for monitoring resource hotspots (mines, monster camps, loot chests). When trust drops because a spot is over-farmed or abused, these controllers automatically adjust cooldowns, spawn rates or access rules to stabilise shared fields without changing formal win conditions.

\item \textbf{Believable cooperative NPCs}: NPC teammates or opponents maintain a trust state toward nearby players. Selfish behaviour (e.g., always grabbing loot) lowers trust and makes NPCs more defensive or greedy, while cooperative play increases trust and thereby also resource sharing and mutual support.

\item \textbf{Pre-launch balancing sandbox}: Large-scale simulations with ETL-equipped agents acting as stand-ins for human players let designers test level compositions and resource layouts  
before release, revealing levels that suffer from over-exploitation
or permanent depletion.

\item \textbf{Harvesting and shared-field management}: ETL-equipped harvesting bots learn to back off stressed resource nodes and spread out over the map instead of over-farming a few hotspots, providing a drop-in micro-management controller for mining or farming systems that improves overall yield without changing the original agent logic.
\end{compactitem}

\section{Related Work}
\label{sec:related-work}

In domains such as large-scale multiplayer games (e.g., StarCraft or Dota), agents developed by different organisations may interact without shared rewards or coordinated training objectives, making centralised training assumptions difficult to satisfy~\cite{OpenAI5, samvelyan2019smac}. More broadly, when cooperative incentives are undefined or structurally misaligned with individual interests~\cite{zhou2023centralized}, it becomes challenging to apply standard cooperative MARL techniques, yet sustaining fair and engaging play still requires some form of emergent coordination.
Promoting cooperation in MARL has therefore been widely studied, especially in environments simulating social dilemmas~\cite{du2023review, wang2022cooperative}. Existing methods generally rely on structural assumptions, such as shared incentives, persistent agent identities, or explicit communication channels, to facilitate cooperation~\cite{cai2022cooperative, wang2023cooperative}. However, these assumptions often do not hold in decentralised or competitive real-world settings, where agents act independently, resource access is unequal, and cooperative behaviour is neither rewarded nor expected~\cite{mushtaq2023multi, canese2021multi}.


A common strategy for inducing cooperation is to modify the reward function or employ centralised training with decentralised execution (CTDE)~\cite{amato2024introduction}. Algorithms such as QMIX~\cite{rashid2018qmix} and MADDPG~\cite{lowe2020multiagentactorcriticmixedcooperativecompetitive} assume global observability during training and often rely on reward shaping to align policies with a predefined cooperative objective~\cite{saifullah2024multi}. While effective in many benchmark environments, these approaches become less applicable when cooperative incentives are undefined or structurally misaligned with individual interests, as in open-ended game worlds with competing stakeholders~\cite{zhou2023centralized}.

Another class of methods enhances cooperation through communication, enabling agents to share observations, intentions, or learned features~\cite{zhu2024survey, hausknecht2016cooperation, singh2018learning}. Models such as CommNet~\cite{sukhbaatar2016learningmultiagentcommunicationbackpropagation}, DIAL~\cite{foerster2016learningcommunicatedeepmultiagent}, and TarMAC~\cite{das2019tarmac} show improved coordination in cooperative tasks where agents interact repeatedly and build mutual policies over time. These methods often assume fixed agent identities and reliable communication protocols. However, in many decentralised environments such as peer-to-peer swarms, anonymised online platforms, or ad-hoc game lobbies, interactions are non-repeating and communication is limited or entirely absent~\cite{khan2023communication}. In such cases, coordination must rely on behavioural inference rather than explicit signals.


Traditional trust-based MARL approaches typically rely on persistent agent identities and repeated interactions to estimate and use trust, often through explicit scores, partner selection, or reputation systems~\cite{fung2024trust, haoran2021multi}. However, such assumptions are unrealistic in many real-world and game settings where agents are anonymous and interactions are non-repeating. In contrast, our algorithm adopts an implicit trust mechanism that requires no communication or identity tracking. Agents infer trust from environmental signals, such as the remaining resource received, rather than direct observation of others. This form of decentralised, outcome-driven trust influences learning dynamics rather than partner choice, enabling cooperation even in anonymous, structurally competitive environments.

\section{Emergent Trust Learning}
\label{sec:etl}

Emergent Trust Learning (ETL) is a lightweight, trust-based control algorithm for decentralised MARL. Rather than replacing an underlying learner, ETL can be attached to standard value- or policy-based agents as an additional state variable and update rule. Each agent interacts with its environment through repeated episodes and receives only its own scalar reward signal. At every timestep, the agent observes a state that may include local environmental features and internal variables (e.g., current payoff or risk level), together with a compact summary of its current trust in others. Based on this augmented state and on stored experience, ETL selects an action, receives an individual reward, and updates both its behavioural policy and trust estimates. 

ETL is easy to plug into existing MARL setups.
\begin{itemize}
    \item it requires \emph{no} explicit communication, shared rewards, or persistent agent identities;
    \item it only uses locally observable information and the agent's own rewards;
    \item it adds only a few scalar variables per agent (trust and simple statistics) and therefore incurs negligible computational and communication overhead.
\end{itemize}

The algorithm 
consists of three interacting components, detailed in the following subsections: \textit{memory} separates short-term and long-term experience to stabilise learning while remaining responsive; \textit{trust} translates locally observed outcomes into expectations about others' behaviour without explicit messages or identity tracking; \textit{exploration} adapts stochasticity based on trust stability and recent performance. This structure applies across all three environments, making ETL an environment-agnostic algorithm.

\subsection{Memory Mechanism}
\begin{figure}[tbh]
    \centering
    \resizebox{0.9\linewidth}{!}{ 
    \begin{tikzpicture}[
        block/.style={rectangle, draw=black, thick, rounded corners=5pt,
            minimum width=3.2cm, minimum height=1.2cm, text centered, fill=blue!5, font=\footnotesize},
        process/.style={rectangle, draw=black, thick, rounded corners=5pt,
            minimum width=3cm, minimum height=1cm, text centered, font=\footnotesize},
        arrow/.style={->, thick, >=stealth}
    ]
    
    \node[block] (input) at (0,2.0) {
        \begin{tabular}{c}
        Experience Input\\
        $(s_t, a_t, r_t)$
        \end{tabular}
    };
    
    \node[block] (stm) at (-1.8,0.6) {
        \begin{tabular}{c}
        Short-term Memory\\
        Recent Interactions\\
        Quick Adjustment
        \end{tabular}
    };
    
    \node[block] (ltm) at (1.8,0.6) {
        \begin{tabular}{c}
        Long-term Memory\\
        Historical Experience\\
        Strategy Guidance
        \end{tabular}
    };
    
    \node[process] (perf) at (-1.8,-0.8) {
        \begin{tabular}{c}
        Recent Performance\\
        Avg Reward: $\bar{r}_t$\\
        Task Success Signals
        \end{tabular}
    };
    
    \node[process] (pattern) at (1.8,-0.8) {
        \begin{tabular}{c}
        Experience Summary\\
        Patterns of Successful\\
        Actions in Context
        \end{tabular}
    };
    
    \node[block] (decision) at (0,-2.4) {
        \begin{tabular}{c}
        Action Decision\\
        $a_t = f(s_t, \text{Experience})$
        \end{tabular}
    };

    \draw[arrow] (input) -- (stm);
    \draw[arrow] (input) -- (ltm);
    
    \draw[arrow] (stm) -- (perf);
    \draw[arrow] (ltm) -- (pattern);
    
    \draw[arrow] (perf) -- (-0.9,-1.7) -- (decision);
    \draw[arrow] (pattern) -- (0.9,-1.7) -- (decision);
    
    \draw[arrow] (decision) to[out=180,in=-90] (-4.0,-1.0) to[out=90,in=210]
        node[left,xshift=-2pt] {\small Update} (stm);
    \draw[arrow] (decision) to[out=0,in=-90] (4.0,-1.0) to[out=90,in=-30]
        node[right,xshift=2pt] {\small Learn} (ltm);
    
    \end{tikzpicture}
    } %
    \caption{Schematic of the memory mechanism in ETL, highlighting how short-term and long-term experience buffers are updated from state-action-reward tuples and jointly drive subsequent action decisions.}
    \label{fig:etl-memory}
\end{figure}
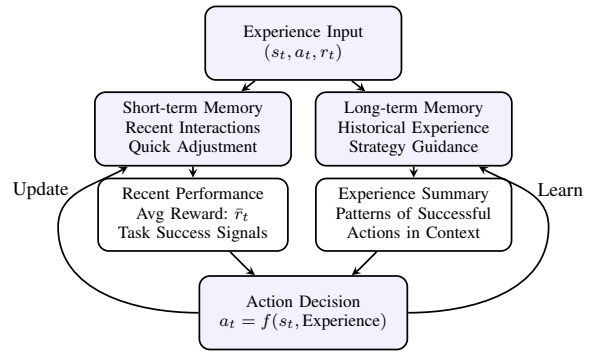

Memory plays a central role in ETL, enabling agents to adapt to environmental fluctuations and learn sustainable strategies. Each agent maintains a short-term memory (STM) for recent experiences and a long-term memory (LTM) for cumulative knowledge. This structure does not assume any particular task: STM tracks how the most recent interaction pattern evolve, while LTM aggregates which behaviours have been successful across many episodes, as shown in Figure~\ref{fig:etl-memory}.

Formally, the combined memory structure is defined as:
\begin{equation}
\mathcal{M}_s = \{(s_t, a_t, r_t)\}_{t=T-N_s}^{T}, \quad
\mathcal{M}_l = \{(s_t, a_t, r_t)\}_{t=0}^{T}
\end{equation}
where each $(s_t, a_t, r_t)$ represents a state-action-reward tuple. STM supports rapid response by tracking the most recent $N_s$ steps, whereas LTM aggregates experience over all $T$ timesteps to inform trust development and policy refinement.

Together, the two memory buffers allow agents to reinforce actions that keep the system in favourable regimes and discard patterns that lead to instability, regardless of the concrete domain.

\subsection{Trust Mechanism}
\begin{figure}[tbp]
    \centering
    \resizebox{0.9\linewidth}{!}{ 
    \begin{tikzpicture}[
        block/.style={rectangle, draw=black, thick, rounded corners=5pt,
            minimum width=3.2cm, minimum height=1.2cm, text centered, fill=green!5, font=\footnotesize},
        process/.style={rectangle, draw=black, thick, rounded corners=5pt,
            minimum width=3cm, minimum height=1cm, text centered, font=\footnotesize},
        arrow/.style={->, thick, >=stealth}
    ]
    
    \node[block] (state) at (0,2.6) {
        \begin{tabular}{c}
        Neighbour Behaviour\\
        Cooperation Signal: $s_t$
        \end{tabular}
    };
    
    \node[block] (current) at (-1.8,1.0) {
        \begin{tabular}{c}
        Current Trust\\
        $T_{ij}^t \in [-1,1]$
        \end{tabular}
    };
    
    \node[block] (history) at (1.8,1.0) {
        \begin{tabular}{c}
        Trust History\\
        Interaction Statistics
        \end{tabular}
    };
    
    \node[process] (update) at (1.0,-0.6) { 
        \begin{tabular}{c}
        Trust Update\\
        $\Delta T_{ij} = f(s_t, d_t)$
        \end{tabular}
    };
    
    \node[block] (final) at (1.0,-2.2) { 
        \begin{tabular}{c}
        Final Trust State\\
        Guides Future Decisions
        \end{tabular}
    };
    
    \node[process] (alpha) at (-2.4,-0.6) { 
        \begin{tabular}{c}
        Learning Rate\\
        Based on Trust Stability
        \end{tabular}
    };
    
    \draw[arrow] (state) -- (-0.9,1.8) -- (current);
    \draw[arrow] (state) -- (0.9,1.8) -- (history);
    
    \draw[arrow] (current) -- (alpha);
    \draw[arrow] (history) -- (update);
    \draw[arrow] (alpha) -- (update);
    
    \draw[arrow] (update) -- (final);
    
    \draw[arrow] (final) to[out=180,in=-90] (-4.5,0.0) to[out=90,in=210]
        node[above left, xshift=-1pt, yshift=0pt] {\small Update} (current);
    \draw[arrow] (final) to[out=0,in=-90] (4.2,0.0) to[out=90,in=-30]
        node[above right, xshift=1pt, yshift=0pt] {\small Record} (history);
    
    \end{tikzpicture}
    }
    \caption{Overview of the trust mechanism in ETL: locally observed cooperation and dissatisfaction signals update a scalar trust state, which is smoothed over time and fed back into decision making without requiring explicit communication or identities.}
    \label{fig:etl-trust}
\end{figure}
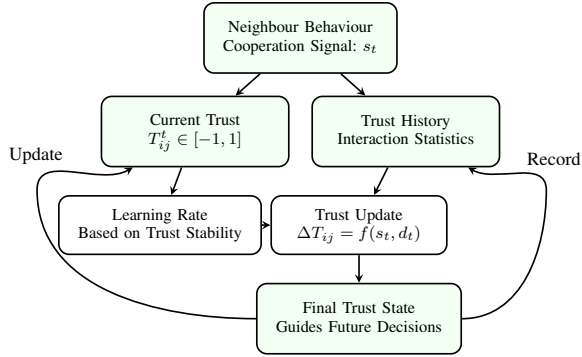

Trust plays a critical role in regulating cooperation within the ETL algorithm. Each agent maintains a dynamic trust value towards its interaction context or an abstract partner index, denoted as $T_{ij}^t$. This value is updated at each timestep based on observed outcomes that reflect how supportive or harmful recent interactions are. Importantly, ETL does not require explicit identity tracking or messages from other agents: trust is inferred solely from locally observable signals and the agent's own reward, as shown in Figure~\ref{fig:etl-trust}.

We define the trust update rule as:
\begin{align}
\Delta T_{ij} &= \alpha \, s_i^t 
              - \beta \, d_i^t 
              + \gamma \left( \bar{T}_{ij}^t - T_{ij}^t \right), \notag \\
T_{ij}^{t+1} &= \mathrm{clip}\left( T_{ij}^t + \Delta T_{ij}, -1, 1 \right)
\end{align}
Here, $s_i^t \in [0, 1]$ is a bounded \emph{support} signal that measures how favourable the current outcome is for agent $i$, while $d_i^t \ge 0$ is a \emph{dissatisfaction} signal that captures perceived harm, risk, or unfairness. The term $\bar{T}_{ij}^t$ is a smoothed long-term trust value. Parameters $\alpha$, $\beta$, and $\gamma$ control responsiveness to positive outcomes, sensitivity to negative outcomes, and the influence of historical trust.

These signals can be instantiated in many ways as long as they are derived from local observations. For example, $s_i^t$ may measure how much of a shared resource remains available nearby, how often joint success conditions are met, or whether payoffs are above a safety threshold; $d_i^t$ may reflect damage received, local resource collapse, or repeated exploitation in a matrix game. In our experiments, we map concrete domain feedback (e.g., remaining food, local resource levels, payoffs in the IPD) onto this pair $(s_i^t, d_i^t)$, so that the same trust update rule can shape behaviour across 
different environments.

\subsection{Exploration Mechanism}
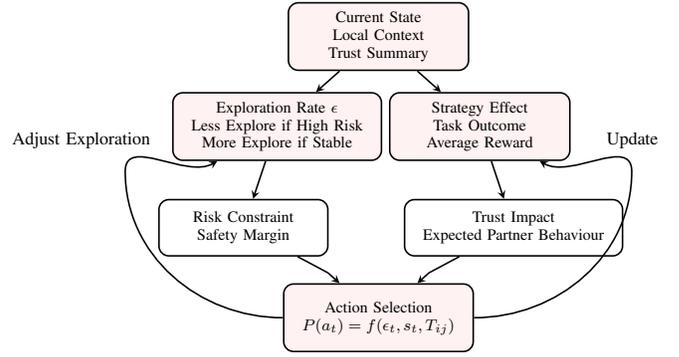
\begin{figure}[htbp]
    \centering
    \resizebox{1.01\linewidth}{!}{
    \begin{tikzpicture}[
        block/.style={rectangle, draw=black, thick, rounded corners=5pt,
            minimum width=3.2cm, minimum height=1.2cm, text centered, fill=red!5, font=\footnotesize},
        process/.style={rectangle, draw=black, thick, rounded corners=5pt,
            minimum width=3cm, minimum height=1cm, text centered, font=\footnotesize},
        arrow/.style={->, thick, >=stealth}
    ]
    
    \node[block] (state) at (0,2.6) {
        \begin{tabular}{c}
        Current State\\
        Local Context\\
        Trust Summary
        \end{tabular}
    };
    
    \node[block] (epsilon) at (-1.8,1.0) {
        \begin{tabular}{c}
        Exploration Rate $\epsilon$\\
        Less Explore if High Risk\\
        More Explore if Stable
        \end{tabular}
    };
    
    \node[block] (perf) at (1.8,1.0) {
        \begin{tabular}{c}
        Strategy Effect\\
        Task Outcome\\
        Average Reward
        \end{tabular}
    };
    
    \node[process] (hunger) at (-2.4,-0.8) {
        \begin{tabular}{c}
        Risk Constraint\\
        Safety Margin
        \end{tabular}
    };
    
    \node[process] (trust) at (2.4,-0.8) {
        \begin{tabular}{c}
        Trust Impact\\
        Expected Partner Behaviour
        \end{tabular}
    };
    
    \node[block] (action) at (0,-2.4) {
        \begin{tabular}{c}
        Action Selection\\
        $P(a_t) = f(\epsilon_t, s_t, T_{ij})$
        \end{tabular}
    };
    
    \draw[arrow] (state) -- (-0.9,1.8) -- (epsilon);
    \draw[arrow] (state) -- (0.9,1.8) -- (perf);
    
    \draw[arrow] (epsilon) -- (hunger);
    \draw[arrow] (perf) -- (trust);
    
    \draw[arrow] (hunger) -- (-0.9,-1.6) -- (action);
    \draw[arrow] (trust) -- (0.9,-1.6) -- (action);
    
    \draw[arrow] (action) to[out=180,in=-90] (-4.5,0.2) to[out=90,in=210]
        node[above left, xshift=-1pt, yshift=2pt] {\small Adjust Exploration} (epsilon);
    \draw[arrow] (action) to[out=0,in=-90] (4.5,0.2) to[out=90,in=-30]
        node[above right, xshift=1pt, yshift=2pt] {\small Update} (perf);
    
    \end{tikzpicture}
    }
    \caption{Exploration mechanism in ETL, where the exploration rate is adapted based on trust stability, perceived risk and recent performance, so that agents explore more under uncertainty and exploit more in stable, trusted regimes.}
    \label{fig:etl-exploration}
\end{figure}

Exploration in ETL is dynamically adjusted based on both trust stability and recent agent performance, enabling a balance between adaptability and convergence, as shown in Figure~\ref{fig:etl-exploration}. At each time step, the exploration rate $\epsilon_t$ is updated as:
\begin{equation}
\epsilon_{t+1} = 
\begin{cases} 
\min(0.9, \epsilon_t \times 1.1) & \text{if } \sigma_T > \theta_T, \\
\max(\epsilon_{\min}, \epsilon_t \times 0.995) & \text{otherwise},
\end{cases}
\end{equation}
where $\sigma_T$ denotes trust variance and $\theta_T$ is the stability threshold. Higher variance increases exploration to encourage policy diversification, while lower variance favours exploitation of learned strategies.

Action selection is further shaped by trust and by how stable the current situation appears to the agent: 
\begin{equation}
P(a_i^t | s_t) = 
\begin{cases} 
1 - \epsilon_t & \text{if } T_{ij} > 0.5 \text{ and } q_i^t > \theta_Q, \\
\epsilon_t & \text{otherwise}.
\end{cases}
\end{equation}
Here, $q_i^t \in [0, 1]$ is a stability score that summarises how safe or reliable the current situation is for agent $i$, and $\theta_Q$ is a threshold that defines when the agent should prefer exploitation over exploration. The score $q_i^t$ can be constructed from any combination of local signals, such as recent reward variability, resource availability, or frequency of successful joint outcomes. This rule ensures that agents exploit cooperative actions when trust is high and the situation is stable, while defaulting to exploration under uncertainty, with a small stochastic component retained to prevent premature convergence to suboptimal behaviours.

\section{Experiments and Results}
\label{sec:experiments}
We evaluate ETL across three environments: a grid-based harvesting game, a hierarchical Tower social dilemma, and the Iterated Prisoner’s Dilemma. Together, they let us test whether the ETL can stabilise shared resources, sustain cooperation under asymmetric roles, and remain effective in a compact game-theoretic setting.
\subsection{Grid-based Resource Environment}

Our first environment is a simple grid-based resource world. A fixed number of agents move on a two-dimensional grid and can harvest from nearby resource tiles; harvested tiles temporarily enter a cooldown state during which they yield little or no reward. Each agent is rewarded purely for its \emph{own} collected resources; any cooperative pattern has to emerge from how agents adapts, not from global rewards that directly pay for cooperation.

\begin{figure}[t]
    \centering
    \includegraphics[width=0.9\linewidth]{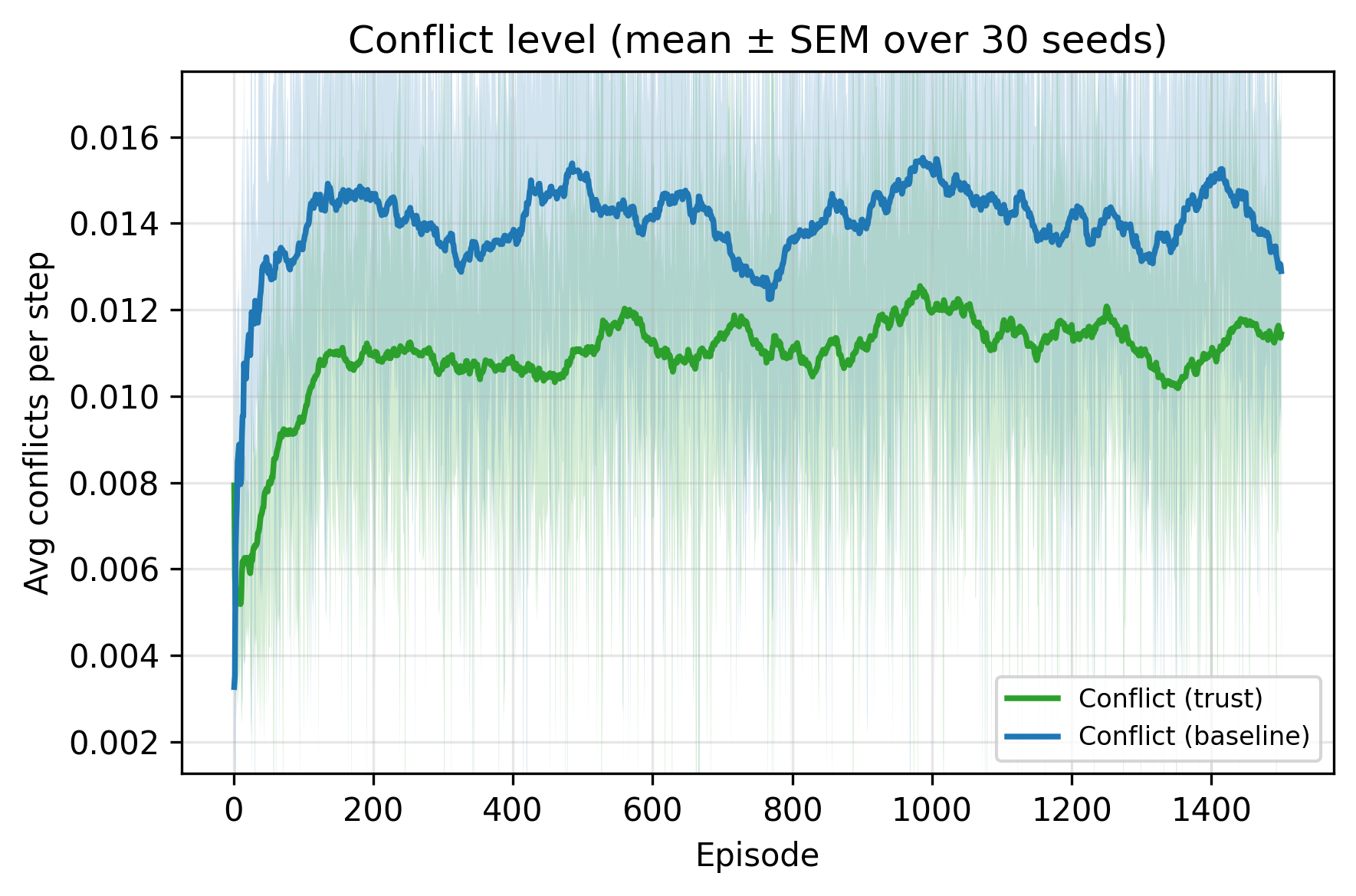}
    \caption{ETL reduces long-run conflict intensity compared to a trust-free baseline, yielding substantially fewer low-value clashes over shared resources across 30 independent runs.}
    \label{conflict}
\end{figure}
We first look at how the trust-based ETL algorithm changes interaction patterns compared to a baseline that uses the same underlying learner without trust. Figure~\ref{conflict} shows the average conflict level, measured as the number of conflicts per step, averaged over 30 independent runs. Across training, ETL consistently produces fewer conflicts than the baseline: agents still compete over resources, but they avoid the most destructive clash patterns that arise when everyone pushes into the same stressed regions at once.

\begin{figure}[b]
    \centering
    \includegraphics[width=0.9\linewidth]{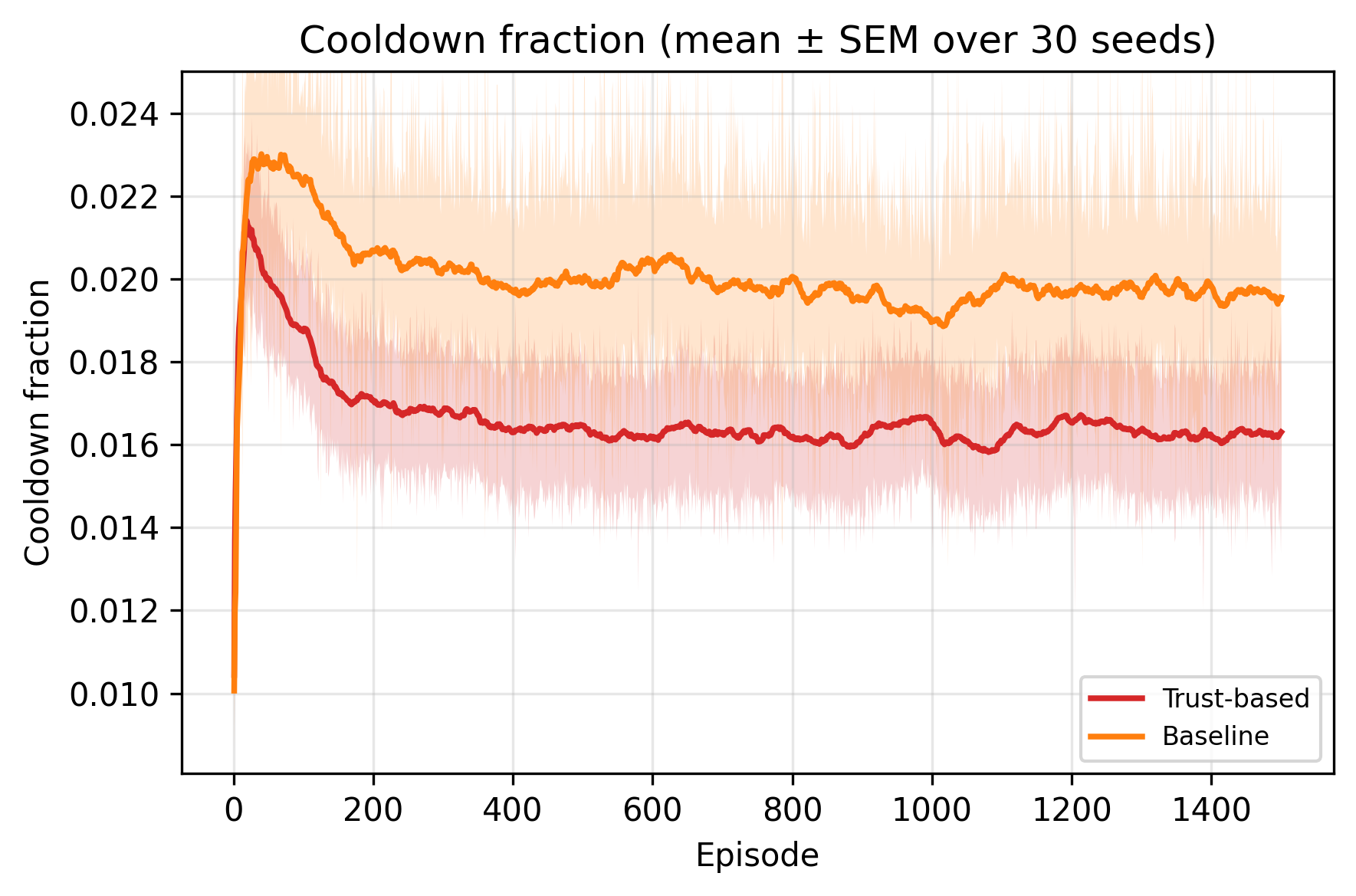}
    \caption{ETL stabilises the environment by keeping the fraction of tiles in cooldown consistently lower than a trust-free baseline, indicating less persistent over-exploitation of shared resources.}
    \label{cooldown}
\end{figure}

The same trend appears in the state of the environment. Figure~\ref{cooldown} reports the fraction of tiles that are in cooldown at each episode. With trust-based ETL, the cooldown fraction is noticeably lower over time: agents learn to back off when the environment shows signs of stress, so that large regions do not remain locked in an unproductive state. In other words, ETL keeps the underlying resource landscape in a healthier regime while optimising the same individual reward signal.

\begin{figure}[tb]
    \centering
    \includegraphics[width=0.9\linewidth]{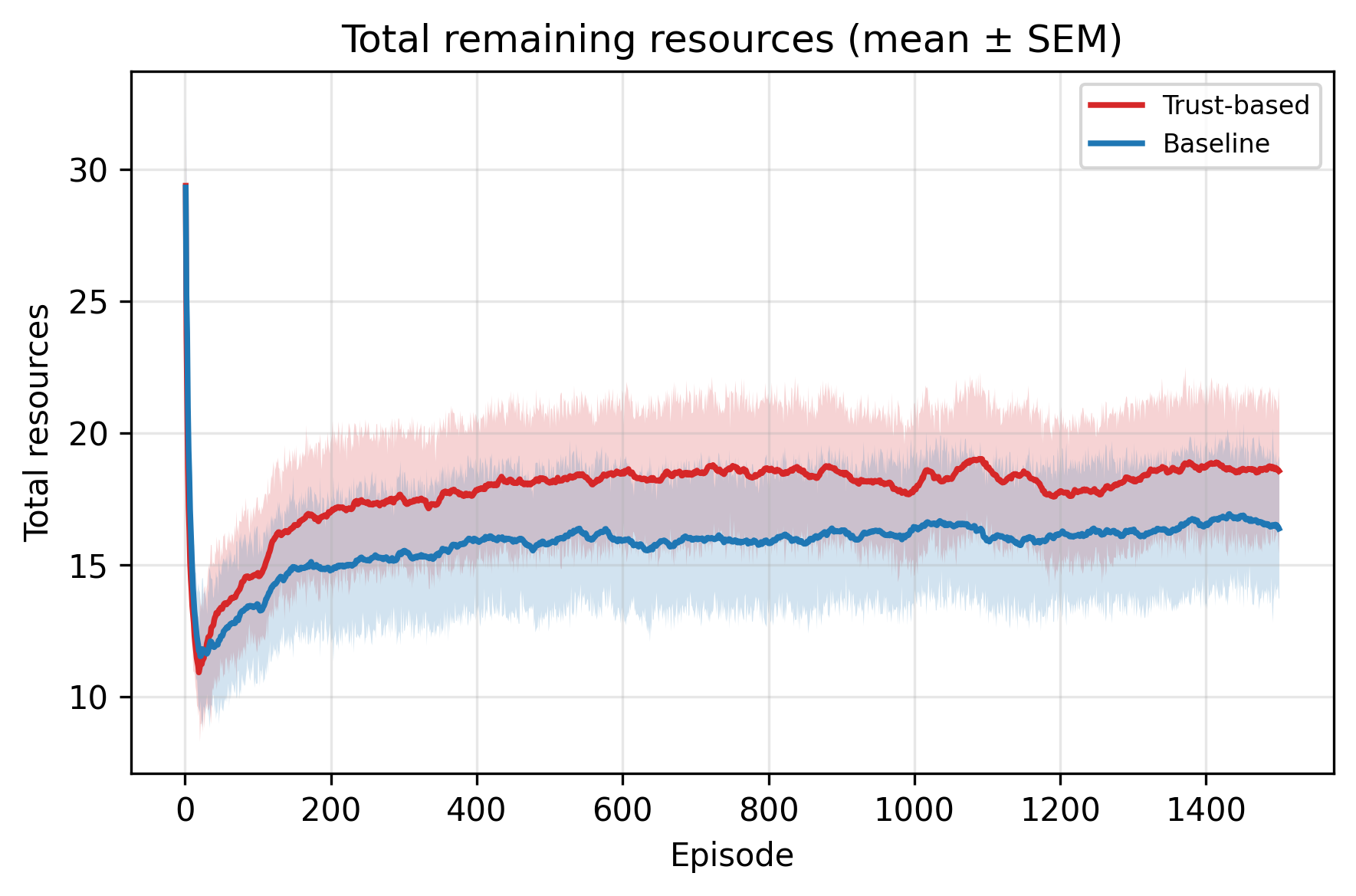}
    \caption{Total remaining resources at the end of each episode in the grid environment (mean over 30 seeds); runs with ETL leave substantially more resources on the map than the baseline while optimising the same individual reward.}
    \label{fig:grid-resources}
\end{figure}

To measure longer-term resource use, Figure~\ref{fig:grid-resources} plots the total amount of resource remaining at the end of each episode. Baseline runs tend to leave the grid more depleted, whereas trust-based ETL consistently preserves more resources for future episodes. This means that ETL agents' learned behaviour is less destructive to the shared environment, and they obtain strong individual returns without exhausting the map.

From a game perspective, these metrics correspond to how often agents fight over the same spot, how much of the map is temporarily ``burned out'' and uninteresting, and how playable the world remains over longer horizons. ETL improves all three while never receiving an explicit reward for cooperation, suggesting that it can act as a lightweight governance algorithm for shared spaces.

\subsection{Tower Environment}

\begin{wrapfigure}{l}{0.4\linewidth} 
    \centering
    \vspace{-10pt} 
    \includegraphics[width=\linewidth]{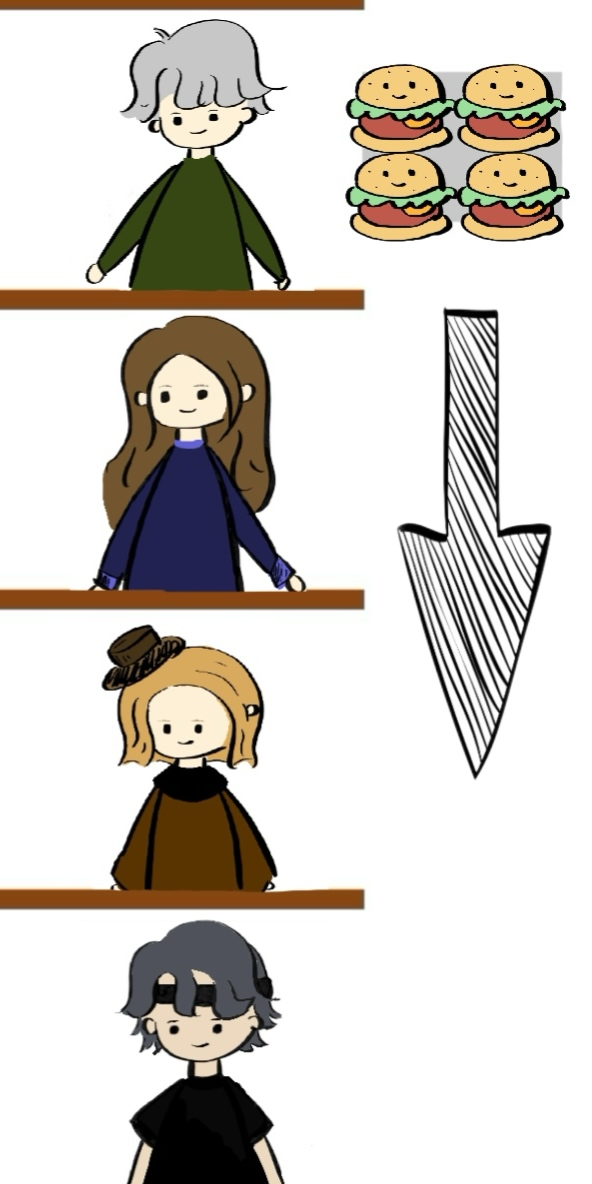}
    \caption{Initial state of the Tower environment. The food platform starts at the top of the tower, and agents are randomly assigned to floors. }
    \label{fig:tower-init}
\end{wrapfigure}

This experiment draws inspiration from the movie ``The Platform''~\cite{theplatform2019}. 
We implemented a simplified visual representation of the tower environment in Pygame. As shown in Figure~\ref{fig:tower-init}, this version uses hand-drawn cartoon illustrations to simulate the resource allocation dynamics.

We design a four-story tower environment with one agent assigned to each floor. The sole resource is food, carried by a platform starting at the top floor and descending one level at a time. The platform starts with 4 food units, just enough for all four agents to survive if each takes exactly 1 unit. When the platform arrives at an agent's floor, they may consume 0, 1, or 2 units. Unused food descends to the next floor, inducing interdependence and social dilemmas.

An agent's hunger level increases linearly over time, and if it reaches the predefined maximum hunger threshold \( H_{\text{max}} \), the agent is considered ``dead". The dynamic hunger update for each agent is governed by the following equation:
\begin{equation}
H_i(t+1) = \min\left(H_{\text{max}}, H_i(t) + \Delta H - \kappa A_i\right)
\label{eq:hunger_update}
\end{equation}
where \( H_i(t) \) is the hunger level of agent \( i \) at time \( t \), \( \Delta H \) is the natural increase in hunger per round, \( \kappa \) is the reduction in hunger per unit of food consumed, and \( A_i \) is the agent's food consumption.

At the end of each round, the positions of all agents are randomly reassigned to different floors, introducing further uncertainty into the environment.

Agents receive rewards based on food consumption, with a penalty for starvation:
   \begin{equation}
   R_i =
   \begin{cases}
   A_i, & \text{if } H_i(t) < H_{\text{max}} \\
   -1, & \text{if } H_i(t) \geq H_{\text{max}}
   \end{cases}
   \label{eq:reward_piecewise}
   \end{equation}
   where \( R_i \) is the reward for agent \( i \), \( A_i \in \{0, 1, 2\} \) is its food consumption, and \( H_i(t) \) is its current hunger level.

This reward structure is entirely short-term oriented, with rewards closely tied to immediate consumption actions. As a result, agents are naturally incentivised to consume as much food as possible to maximise their immediate returns, often at the expense of agents on lower floors, while the absence of explicit cooperation rewards and the random reassignment of floor positions together make it difficult to develop stable collaborative strategies.
\vspace*{1ex}

Figure~\ref{fig:tower-overview} illustrates two typical scenarios. The left side illustrates agent starvation due to upper-floor greed.
The right side shows a situation where all agents survive when agents learn to moderate their consumption to exactly 1 food unit each.

We evaluate ETL's ability to foster cooperation without explicit cooperative rewards by comparing it to traditional MARL algorithms and testing robustness under different initial conditions.

\subsubsection{Comparison with Baselines}
As comparison, we re-implemented standard tabular Q-Learning and Monte Carlo control following \cite{sutton1998reinforcement}, with $\epsilon = 0.1$, and for Q-Learning, $\alpha = 0.1$, $\gamma = 0.95$. Both agents use epsilon-greedy policies over discrete state-action spaces.

\begin{figure}[t]
    \centering
    \includegraphics[width=0.9\linewidth]{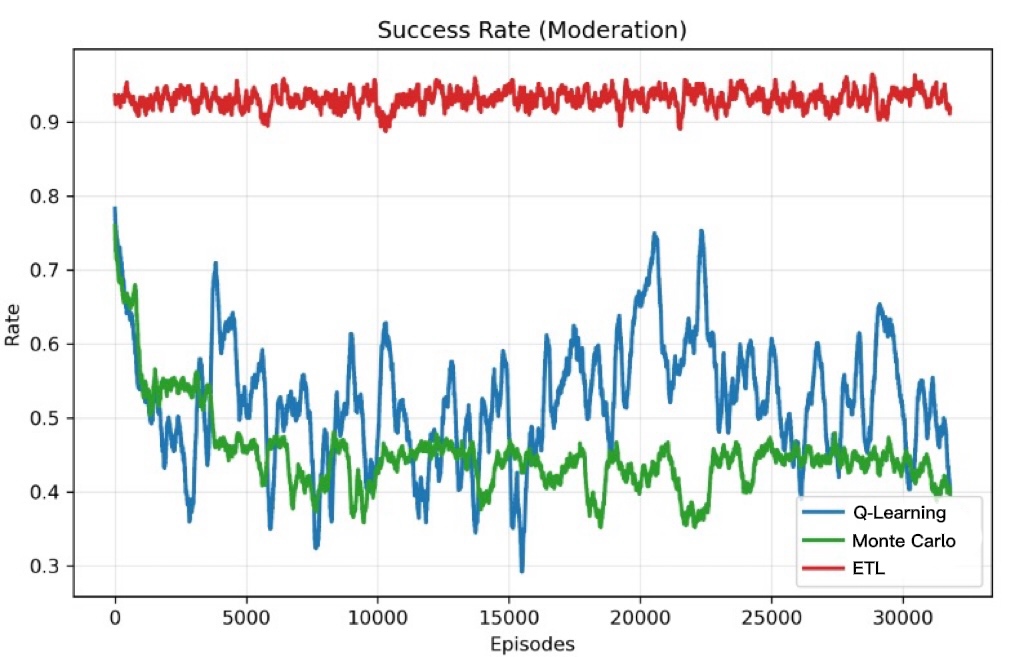}
    \caption{Success rate in the Tower environment over training episodes, comparing ETL to tabular Q-Learning and Monte Carlo control. Only ETL achieves and maintains a high success rate 
    without explicit cooperative rewards.}
    \vspace{2em}
    \label{fig:success-comparison}
\end{figure}

Figure~\ref{fig:success-comparison} shows that both Q-Learning and MC fail to achieve stable cooperation in the Tower Environment, with success rates oscillating or stagnating due to the absence of explicit cooperative rewards. In contrast, ETL maintains a near-constant success rate above 90\%.

Interestingly, the success curve of ETL remains flat and high throughout training. This may indicate that once trust equilibrium is reached, agents do not deviate, suggesting fast convergence under favourable initialisation. 

\subsubsection{Robustness to Greedy Initialisation}
To test the robustness of ETL under unfavourable conditions, we simulate extreme early selfishness by forcing all agents to follow a greedy strategy (consuming 2 food units) for a fixed number of initial episodes. The objective is to determine whether cooperation can still emerge after such behaviour and how long it takes for the system to recover.

\begin{figure}[b]
    \centering
    \includegraphics[width=0.85\linewidth]{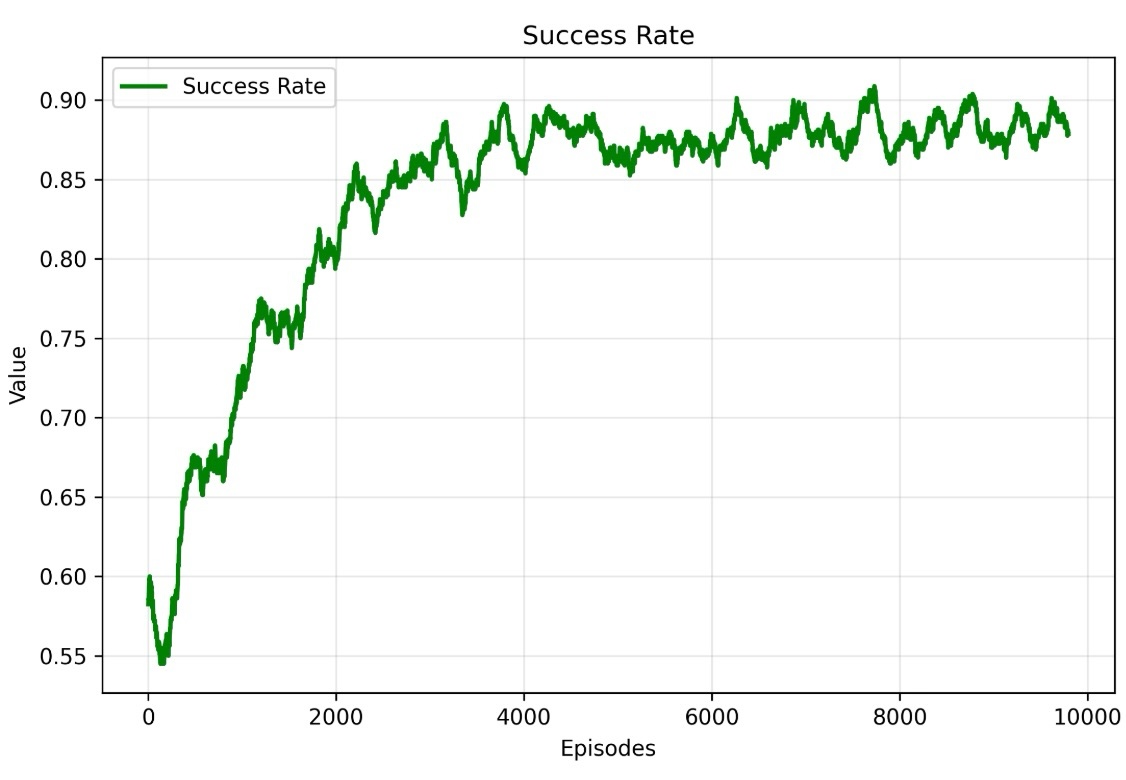}
    \caption{Recovery of ETL in the Tower environment after 200 episodes of forced greedy behaviour: the success rate starts near zero during enforced greed but climbs back above 90\% once learning resumes, demonstrating robustness to unfavourable initial conditions.}
    \vspace{1em}
    \label{fig:success-recovery-greedy}
\end{figure}

As shown in Figure~\ref{fig:success-recovery-greedy}, during the first 200 episodes of enforced greediness, the success rate remains near zero, indicating complete failure of cooperation. Once the learning phase begins, ETL agents quickly recover: success rates climb rapidly, surpassing 90\% after approximately 3000 episodes. This demonstrates the algorithm's strong adaptive capability, even after a period of extreme non-cooperative behavior.

\begin{table}[tbh]
    \centering
    \begin{tabular}{c|c|c}
        \hline
        \textbf{Player A} & \textbf{Player B} & \textbf{Payoff (A, B)} \\
        \hline
        Cooperate (C) & Cooperate (C) & (3, 3) \\
        Cooperate (C) & Defect (D)    & (0, 5) \\
        Defect (D)    & Cooperate (C) & (5, 0) \\
        Defect (D)    & Defect (D)    & (1, 1) \\
        \hline
    \end{tabular}
    \caption{Payoff matrix of the Iterated Prisoner’s Dilemma used in our experiments, with the standard ordering where unilateral defection yields the highest individual payoff but mutual cooperation dominates mutual defection.}
    \label{tab:payoff_matrix}
\end{table}

This pattern suggests that ETL does not simply memorize early policies but continues to reinterpret outcomes through its trust signal, allowing cooperative behaviour to re-emerge even after an extended phase of enforced defection. In practical terms, this makes the algorithm robust to unfavourable initial conditions or early misalignment in deployed systems.

\subsection{Prisoner’s Dilemma}

The Iterated Prisoner’s Dilemma (IPD) provides an established, game-theoretic testbed for studying how agents balance short-term exploitation against long-term cooperation.
In each round, two players simultaneously choose whether to cooperate (C) or defect (D), and receive payoffs according to the standard Prisoner’s Dilemma matrix in Table~\ref{tab:payoff_matrix}.
Mutual cooperation yields a solid but moderate payoff for both players, unilateral defection grants a one-shot advantage to the defector, and mutual defection traps both sides in a low but safe outcome.

Our goal in this environment is twofold.
First, we want to verify that ETL remains competitively strong when faced with a diverse meta-game of opponents.
Second, and more importantly from a game-design perspective, we want to know whether its internal trust mechanism can be manipulated by opponents that explicitly try to exploit it.
To this end, we consider two groups of strategies.
The baseline set contains classic IPD archetypes: Always Cooperate, Always Defect, Random, and Tit-for-Tat (TFT), which mirrors the opponent’s previous move.

We then introduce two delayed-response strategies that first cooperate or defect for an initial phase before switching behaviour, in order to test whether trust can be gamed or can recover after a rough opening.

We evaluate ETL and all comparison strategies in a fully symmetric round-robin tournament.
For each ordered pair of strategies, we play repeated IPD games of 500 rounds.
Agents are freshly instantiated at the start of each game, so ETL learns only within the current match and does not carry knowledge across opponents.
As our primary scalar performance measure, we use the \emph{success rate}: the proportion of games in which a strategy’s average per-round payoff exceeds the cooperation threshold (greater than $2.5$ in our payoff matrix), i.e., games where the strategy “does well” against its current opponent.

\begin{figure}[t]
    \vspace*{-1.85ex}
    \centering
    \includegraphics[width=\linewidth]{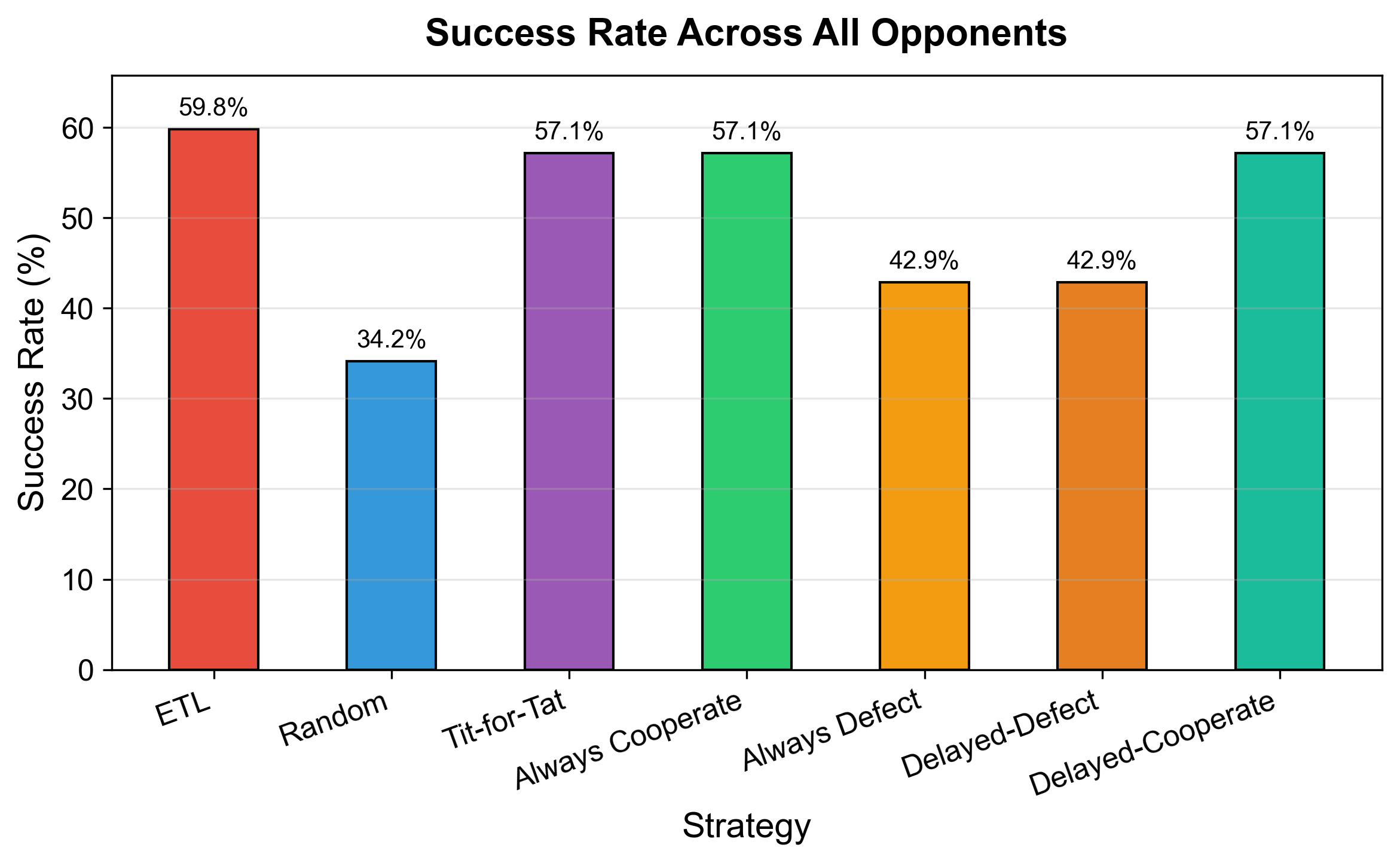}

    \caption{Success rate across all opponents in the Iterated Prisoner’s Dilemma round-robin tournament; ETL outperforms classic strategies such as Tit-for-Tat, Always Cooperate and Always Defect, including against delayed-response adversaries that attempt to exploit its trust mechanism.}
    \label{fig:ipd-winrate-all}
\end{figure}

Figure~\ref{fig:ipd-winrate-all} shows success rates across all opponents.
Each bar aggregates performance over the entire tournament and indicates how often a given strategy manages to secure a strong outcome, regardless of which opponent it faces.

ETL achieves the highest overall success rate among all the strategies, outperforming classical baselines and adversarial delayed-response strategies that attempt to exploit its trust mechanism.
This indicates that the trust mechanism does not simply make the agent ``nice" or easy to exploit.
Against exploitative opponents (Always Defect, Delayed-Defect), ETL initially loses some rounds due to its moderate starting trust, but then reduces cooperation as negative outcomes accumulate, preventing long-term exploitation.
Against cooperative opponents (Tit-for-Tat, Delayed-Cooperate), ETL recognises consistent cooperative behaviour, drives trust upwards, and stabilises interactions near mutual cooperation, with a large share of rewards coming from C-C outcomes rather than unilateral exploitation.
ETL therefore learns to treat early cooperation as a signal rather than a guarantee: it rewards consistent cooperation with stable joint play, but reacts to sudden shifts by gradually withdrawing trust.
From a game context perspective, ETL behaves like a strong but non-toxic player archetype: it wins often in a heterogeneous meta by keeping interactions in healthy, high-cooperation regimes whenever possible, instead of collapsing into low-value defection equilibria.

\section{Discussion}\label{discussion}
Across three qualitatively different environments, we show that a simple, trust-based control algorithm can steer purely self-interested agents towards cooperative regimes without any explicit cooperative rewards. In the grid-based resource world, ETL reduces low-value conflicts and prevents long term resource depletion while optimising the same individual reward as a baseline without trust. In the Tower environment, it sustains near-perfect survival in a hierarchical social dilemma and recovers cooperation even after extended phases of enforced greed. In the IPD, the same mechanism transfers to an abstract matrix game and maintains cooperation with reciprocal opponents while avoiding long term exploitation by strategies that rely heavily on defection.

From a game design perspective, the empirical results suggest several concrete use cases. In grid-like online worlds, trust-based agents can serve as a background controller for resource hotspots: when a mine or monster camp is over farmed, falling trust automatically induces more conservative behaviour or triggers soft balancing (e.g., longer cooldowns or tougher guards) instead of hard coded nerfs. As cooperative NPCs or bots, agents equipped with ETL can adjust their behaviour towards nearby players based on observed sharing or hoarding, becoming more supportive around fair teammates and more defensive around selfish ones. In harvesting-style games, trust-driven policies can stabilise shared fields or forests by discouraging aggressive over exploitation, keeping the environment playable over long sessions without changing core win conditions.

While our study focuses on relatively small populations and stylised environments, the results point to several directions for future work. One avenue is scaling ETL to richer game genres and larger, heterogeneous agent populations with dynamic entry and exit, investigating how trust-based control behaves in open-world or live service settings. Another is combining ETL with deep MARL architectures and centralized training pipelines, using trust as an additional signal rather than a replacement for value-based coordination. Finally, integrating ETL into commercial game development workflows as a general-purpose governance and balancing module for hotspots, harvesting systems, and cooperative NPC behaviour offers an exciting opportunity to test its impact on long term play experience in real player communities.

\section{Conclusion}
\label{sec:conclusion}
We introduced Emergent Trust Learning (ETL), a lightweight trust-based control algorithm that can be plugged into standard value- or policy-based learners 
using only local outcomes and individual rewards. 
Across three qualitatively different environments, 
ETL reliably steers purely self-interested agents toward cooperative regimes. In a grid-based resource world, ETL agents reduce low-value conflicts and long-term depletion while optimizing the same individual reward as a trust-free baseline. In hierarchical Tower social dilemma testbed, ETL sustains near-perfect survival and can recover cooperation even after extended phases of enforced greed, demonstrating that trust dynamics can stabilize play in strongly asymmetric, high-stakes settings. In the Iterated Prisoner’s Dilemma, the algorithm transfers to an abstract matrix game, where it maintains cooperation with reciprocal opponents while remaining robust against exploitative strategies that rely heavily on defection.
Taken together, these results demonstrate that ETL generalizes across domains:
without changing its core update rules, it acts as an effective governance layer for shared resources, a stabilizer for hierarchical social dilemmas, and a strong yet non-toxic player archetype in strategic interactions. Combined with the reusable Tower environment and the envisioned game AI use-cases in Section~\ref{sec:use-cases}, ETL provides a practical, low-overhead building block for designers and researchers who want agents that automatically learn when to share, when to hold back, and how to keep shared worlds playable over time.


\bibliographystyle{IEEEtran}
\bibliography{references}







\end{document}